# Going beyond the one-off: How can STEM engagement programmes with young people have real lasting impact?

Manuscript in preparation for Research for All

## Authors

Martin Archer*, School of Physics and Astronomy, Queen Mary University of London (martin@martinarcher.co.uk, https://orcid.org/0000-0003-1556-4573) *now at Department of Physics, Imperial College London
Jennifer DeWitt, UCL Institute of Education, University College London, London; Independent Research and Evaluation Consultant (jenedewitt@gmail.com, https://orcid.org/0000-0001-8584-2888)
Carol Davenport, NUSTEM, Northumbria University (carol.davenport@northumbria.ac.uk, https://orcid.org/0000-0002-8816-3909)
Olivia Keenan, South East Physics Network (outreach@sepnet.ac.uk)
Lorraine Coghill, Science Outreach, Durham University (l.s.coghill@durham.ac.uk)
Anna Christodoulou*, Department of Physics, Royal Holloway University of London (anna.christodoulou@essex.ac.uk) *now at University of Essex
Samantha Durbin, The Royal Institution (sdurbin@ri.ac.uk)
Heather Campbell, Department of Physics, University of Surrey (h.campbell@surrey.ac.uk)
Lewis Hou, Science Ceilidh (lewis@scienceceilidh.com)

## Abstract

A major focus in the STEM public engagement sector concerns engaging with young people, typically through schools. The aims of these interventions are often to positively affect students' aspirations towards continuing STEM education and ultimately into STEM-related careers. Most schools engagement activities take the form of short one-off interventions that, while able to achieve positive outcomes, are limited in the extent to which they can have lasting impacts on aspirations. In this paper we discuss various different emerging programmes of repeated interventions with young people, assessing what impacts can realistically be expected. Short series of interventions appear also to suffer some limitations in the types of impacts achievable. However, deeper programmes that interact with both young people and those that influence them over significant periods of time (months to years) seem to be more effective in influencing aspirations. We discuss how developing a Theory of Change and considering young people's wider learning ecologies are required in enabling lasting impacts in a range of areas. Finally, we raise several sector-wide challenges to implementing and evaluating these emerging approaches.

## Key Messages

- One-off and short-series of STEM interventions with young people don't appear to have the long-term impacts on aspirations that universities and other practitioners of STEM engagement are often hoping to achieve
- Deeper programmes of engagement are required based around Theories of Change and considering young people's wider learning ecology
- Many sector-wide challenges exist to implementing and evaluating the long-term impacts of such programmes

## Introduction

Engaging schools and young people with Science, Technology, Engineering and Mathematics (STEM) has long been a priority for STEM engagement bodies and practitioners, particularly in the cases of universities/researchers — a survey of UK researchers found that engineering and physical sciences researchers place more importance on engaging school students out of all possible publics than researchers in other areas do (Hamlyn et al., 2015). This is perhaps driven by concerns over the perceived low numbers of young people opting for studies in STEM subjects (e.g. Smith, 2004). Researchers' and practitioners' motives to engage young people in order to encourage them to study a particular subject or even to consider a particular institution, often what are meant by the terms "outreach" and "recruitment" respectively, are typically conflated with public engagement, which in its purest sense is formed of genuine two-way interactions for mutual benefit, and likely affects their approach to engagement practice with schools (Hamlyn et al, 2015). For example, Thorley (2016) found that physicists consider engaging young people as typically transmissive (i.e. one-way) in nature and while they place high levels of importance on a range of content types and messages for young people (knowledge, excitement, relevance, careers information etc.) they unilaterally considered their own role in conveying these topics as less important. Therefore, it appears that in general more critical consideration is needed in recognising audiences' needs and appropriate methods of engagement in order to improve the efficacy of raising STEM aspirations in young people.

STEM engagements with young people have generally taken the form of various one-off experiences, such as a school trip, a show, a speaker, a video or some other interaction. While popular, these are realistically limited in the types of impacts they might be able to have and the likelihood that these impacts will last in the longer term. There is evidence that one-off experiences can in the short-term support content learning (likely just in the form of 'factoids'), as well as lead to affective outcomes, such as interest and inspiration (Bell et al., 2009; DeWitt & Storksdieck, 2008). Likewise, meeting a scientist can certainly provide students with an increased awareness of what a particular career involves or the range of careers that might be available, and even what courses might be required to progress (Woods-Townsend et al.,

2016). Memory research, however, suggests that without further reinforcement of these learning outcomes they will likely be limited to days or weeks (e.g. Murre and Dros, 2017). Furthermore, such one-off experiences are unlikely to have a lasting measurable effect on aspirations. Research into young people's aspirations (cf. L. Archer & DeWitt, 2017, among many others) highlights that aspirations are complex and multifaceted and evolve over time. They are influenced by a range of factors, including experiences at home, in the school and in the wider community, as well as background factors that are interwoven with the way young people experience school, engagements with science, and everything else. They are also closely linked to identity - to what or whom young people can see themselves becoming, and what type of person they (and others) perceive them to be. Consequently, it should not be surprising that one-off experiences, while inspirational and having potential to support a range of outcomes, are unlikely to significantly impact aspirations and educational trajectories. At the same time, because aspirations develop and are maintained (or not) over time, there can be a role for such experiences in providing additional support in their maintenance. Moreover, when considered from the perspective of learning ecologies (discussed in more detail later), any potential impact of a one-off experience can be extended by linking to other experiences that young people may have, both shorter term and longer in duration. In other words, while one-off experiences can be worthwhile experiences, it is important to be realistic about what they can achieve by themselves - and thus to maximise the opportunities they offer and their potential impact by linking to other aspects of young people's lives and experiences.

As we aim to move towards more impactful engagement and deeper learning, we have to carefully consider and evaluate the different types of engagement that we develop. As the sector has moved from the more didactic 'Public Understanding of Science' towards more egalitarian constructivist approaches, we still see the dominance of traditional interventions such as lectures. We are aware of the limitations of the traditional "listen to me" lecture-style event (e.g. Freeman, 2014; Marbach-Ad, 2000; Marbach-Ad, 2001) and in 2013, The Times Higher Education noted over 700 studies determining that lectures were less effective than other teaching strategies (Gibbs, 2013). As such, there are definite moves in undergraduate teaching away from such traditional lecturing with shifts in: the format of lecturing, such as segmenting and including discussion-based approaches (e.g. Iowa State University, 2020); increased use of active learning strategies (e.g. Durham University, 2019); and even the redesigning of learning spaces or "build pedagogy" (Monahan, 2002; Elkington, 2019). However, learners and teachers will often state a preference for lectures over active learning (Deslauriers, 2019). Students feel that they learn more from lectures, even although the evidence suggests otherwise, perhaps partly caused by the additional cognitive effort required by active learning methods (Deslauriers, 2019). Against this backdrop, in trying to develop STEM public engagement projects that have greater scope to develop and support students' aspirations, we are also faced with the additional resources, costs, relationship-building and time required to make such programmes happen.

Nonetheless, several organisations and practitioners have been moving towards engagement programmes of repeated-interventions with the same group of young people in order to maximise the likelihood of impacts on their aspirations towards STEM. This paper provides a landscape review and commentary aimed at STEM engagement practitioners (be they independent or based in a university or institutional setting) on some of the different approaches that have emerged, predominantly in the United Kingdom. It explores their potential benefits and limitations by drawing from evaluation literature as well as social science and educational research/theory. Current challenges faced by the STEM engagement sector and what might be required in order to move forward effectively are also discussed. These topics arose from discussions at a session on repeated interventions at the BIG STEM Communicators Network's BIG Event 2019 (M.O. Archer et al., 2019).

## Short series

The next logical step on from a one-off intervention is to instead deliver a short series of (typically a few) interventions. Here we discuss some different examples of such series delivered within the academic year and their evaluation thus far. We note that other programmes working on interventions across multiple years (though with only one session per academic year) are starting to emerge, but we are not aware of sufficient published evidence of the impacts of these types of initiatives yet.

Activities like summer schools, typically run for late secondary and sixth-form students (aged 15-18), in some sense can be thought of as a short series or "intensive one-off", since they are highly focused down into often just a single week. It appears that summer schools, while enjoyable to the participants, only cause moderate changes to students' likelihood to apply to a selective university in surveys immediately afterwards (e.g. Universify Education, 2018) and many summer school programmes show no change in STEM aspirations from before to after (e.g. Bhattacharyya et al., 2011; MacIver and MacIver, 2015). There is also a critical lack of longitudinal studies at present, as highlighted in a recent review of the higher education widening access sector (Robinson and Salvestrini, 2020). Such studies are necessary in demonstrating whether any increased uptake of higher education actually occurs and whether there is causality in summer schools themselves (and not other factors) leading to improved progression. We also note that summer schools are expensive to deliver and often have severely limited places, making them highly competitive. Even with targeting of participants using widening participation criteria, they are likely selecting predominantly those who are already highly bought into the subject of the summer school. This is an aspect that TASO (https://taso.org.uk/) plan to investigate with randomised control trials in the near future.

'I'm a Scientist, Get Me Out of Here' (https://imascientist.org.uk/) involves an online chat between students (between ages 9-18) and scientists followed by more extended Q&A over the span of typically 2 weeks. They commissioned an evaluation using the framework of science capital to understand what impact it might be having and what might be contributing to that impact (DeWitt, 2019). Perhaps not surprisingly, one of the biggest impacts was on how young people perceived scientists - as normal, regular people, with hobbies, families and interests outside of science. Such perspectives on scientists are similar to those shared by individuals with higher levels of science social capital - who know people (e.g. family, friends' parents) who work in science. While there was limited evidence that this awareness completely changed aspirations of young people, there was an increased willingness among many to consider the possibility of pursuing science further. Of course, the realisation of any longer-term or significant impact on aspirations is dependent on many experiences that may (or may not) happen after. But this increased openness to the possibility of pursuing science, which seems to be influenced by young people realising that scientists are people 'like me', can have a role to play.

A similar shift in perceptions of scientists was found among primary school pupils participating in 'Scientist of the Week', an intervention developed by the NUSTEM team at Northumbria University. This is a five week, teacher-led intervention, in which a new scientist was 'introduced' to the students each week using presentation slides, classroom posters and postcards to take home. In the materials provided to teachers there was a short paragraph describing the work and life of each scientist which included three key attributes of that person, attributes which both contrasted with stereotypical views of scientists (e.g. curious, open-minded, creative) and represented characteristics that young children could imagine themselves possessing and often likely already possessed. In other words, these attributes also communicated that these scientists were 'like me'. Evaluation of this project provided encouraging evidence that young people's perceptions of scientists were shifting - after the intervention pupils were more likely to use non-stereotypical words than stereotypical words when asked to describe a scientist (Shimwell et al., under review).

L. Archer et al. (2014) report on a six-week STEM careers intervention series for Year 9 (13-14 year-old) students at a London girls' school that combined multiple activities that were co-designed and delivered by classroom teachers. The comprehensive evaluation involved before and after surveys, classroom observations, post-intervention student discussion groups, and teacher interviews. While the series did appear to have a positive effect on broadening students' understanding of the range of jobs that science can lead to or be useful for, it did not significantly change students' aspirations or views on science.

Across the South East Physics network (SEPnet, http://www.sepnet.ac.uk) two programmes targeting Year 8 (12-13 year-old) students using multiple interventions have been implemented. The first of these is the Connect Physics programme, a series of three workshops spread out across the academic year which address the following aspects:
1. What is physics? The breadth of the subject and how it all connects together
2. Why do physics? The skills and range of careers open to physicists
3. How do we do physics? The process of the scientific method

The aim of Connect Physics is to raise (or at least maintain) the science capital of children at this age (L. Archer et al., 2015). An evaluation of the pilot of the programme in the academic year 2017-2018 was commissioned externally. The final report (Hope-Stone Research, 2018) showed that the workshops were enjoyed by participants but unfortunately the data was not robust enough to demonstrate any of the aspirational impacts that were hoped for. The key issues were that individual students' responses across the year could not be linked, and there was very little crossover between students answering the initial survey and those that answered the final one. Furthermore, responses could not be separated out by demographic information - so it was not possible to explore whether more or less impact was being had on certain groups. As the programme has become more established, further in-depth evaluation is being conducted (in the academic years 2019-2020 and 2020-2021). This evaluation data will allow linking of individual students' responses, exploration of themes emerging alongside demographic information, and comparisons with control data from Year 8 students at the same schools not taking part in the programme. It is hoped that this will allow a more robust assessment of the impact of this multiple intervention programme. To further support this programme, and to create the opportunity for wider reach and increased likelihood of impact lasting, SEPnet will be creating a teacher Continuing Professional Development (CPD) programme which focuses on science capital, science identity, and the aspirations of students. The aim of this is to further embed the messaging of Connect Physics into the classroom outside of the series of external interventions.

The second programme running across SEPnet (mainly supported by the University of Surrey and Royal Holloway, University of London) is Shattering Stereotypes. Its aim is to explore what role physics Outreach Officers can play in tackling gender stereotyping in schools. The programme was informed by the IOP (2013) Closing Doors report into gender and subject choice and the IOP (2015) Opening Doors guide to good practice in challenging gender stereotypes in schools. Year 8 students are the primary audience, with Year 12 (16-17 year-old) students involved as session facilitators. While students enjoyed the original pilot workshops during the 2016/17 academic year and recognised the relevance of gender stereotyping (see evaluation report of Jeavans and Jenkins, 2018), unfortunately the programme did not result in any significantly changed attitudes in before vs after surveys, either within gender or in gender variations, other than an increased feeling by both genders that physics is difficult. This latter change, however, is likely to be a common outcome over the course of Year 8 anyway (cf. DeWitt et al., 2019) and therefore unlikely to be caused by Shattering Stereotypes. Following the external evaluators' recommendations, the programme was revised in 2017/18 and a second pilot was launched in 2018/19, going on until 2020. Shattering Stereotypes currently consists of three workshops, aiming to raise awareness of what gender stereotypes are, in the context of a student's everyday life and a student's possible career path. The project also aims to empower students so they can identify and challenge situations where they are presented with these stereotypes. The three workshops shaped around these aims, again similarly spread out across the academic year in their delivery, are:
1. Gender Stereotyping & You: Introduction to the concepts of gender stereotyping and how it can have an effect on their lives at home or at school.
2. Gender Stereotyping & Your Career: How gender stereotyping can have an effect on a student's chosen career path, using physics as a case study.
3. Communications Challenge: Students suggest different methods of communicating issues around gender stereotyping to various target groups, including teachers, parents and younger students, giving them the opportunity to take ownership over the project.

The interventions use existing resources from the IOP and the Women in Science and Engineering campaign (i.e. the People Like Me quiz). Schools are also offered options for teachers' involvement, to raise awareness of how gender stereotyping might affect their teaching. Parental engagement is also one of the programme's secondary objectives which has proven challenging in the past, so the second pilot aims to identify new ways of including parents in the discussion. A similar approach to the revised Connect Physics evaluation is being taken with the second pilot of Shattering Stereotypes to allow for more robust analysis. Additionally, SEPnet is broadening the focus of this project to include gender stereotyping and unconscious bias CPD for teachers. A session on this has been developed and trialed and will be offered to teachers (through partner organisations) from Autumn 2020. Additionally this CPD will collaboratively form one module of an Open University OpenLearn course being developed as legacy of Cardiff University's 'Physics Mentoring Project'.

The final example we present here is The Royal Institution's 'Ri Masterclass' programme. These short series of STEM extra-curricular enrichment workshops have been running since 1981. Initially focused on mathematics at secondary level (age 13-14), it has expanded to include both primary and older age-groups as well as engineering and computer science. Teachers select students who they feel should attend. Though initially aimed at the most able, Masterclass organisers now encourage teachers to choose those they feel would benefit most. While only a few places per school are available, teachers are also encouraged to attend and share their experiences back at school. In 2019, 164 Masterclass series ran across the UK for 6276 students, with over 100,000 students having attended since 1981. The aims of the programme (Royal Institution, 2018) have not significantly changed since its inception, and there is a correlation with some of the areas highlighted in science capital research as important in influencing students' aspirations (L. Archer et al., 2015). The main aims are:
- To improve attitudes towards and understanding of these subjects, their applications and their relevance in the wider world
- To allow participants to explore topics interactively and in depth outside of what they would see in the classroom over an extended period of engagement
- To enable participants to meet a range of subject experts and enthusiasts, showcasing a variety of careers
- To enable participants to meet like-minded peers from different schools across the local area

Evaluation has revealed some increases in confidence and positive influences on reported levels of interest in the subjects and future subject choices (Royal Institution, 2018; Barmby et.al., 2008). However, these surveys were a small sample of the students involved and it has been difficult to perform meaningful longitudinal evaluation. In addition, due to the franchise model of the programme, the quality of provision can vary across the country. A significant number of students choose to attend follow-up events, however, which is positive (Royal Institution, 2018).

As with one-off interventions, a short series of engagements does appear to have some impact. However, practitioners need to be realistic as to what types of impacts are achievable from these programmes.

## Deeper programmes

Research suggests that programmes with longer engagement can lead to considerable increases in students' aspirations as well as other outcomes (Robinson and Salvestrini, 2020). These programmes often combine several different approaches in supporting and interacting with students. Here we detail a few different examples.

The NUSTEM primary partnerships are an ongoing collaboration that works with children, teachers and families to support a broadening of aspirations, with a particular focus on STEM careers. There are currently 33 schools in the partnership, with around 12 having been in partnership for over five years. NUSTEM delivers regular activities in the classroom, but also supports teachers to deliver STEM career focused activities of their own through Continuing Professional Development (CPD) and resource development (e.g. the STEM Person of the Week series mentioned previously) as well as working with parents and carers. Over the first three years of the partnership the gender difference in career aspirations between girls and boys decreased significantly, with more children saying that they would be interested in a broader range of careers (Emembolu et al., 2020).

Working in partnership with several organisations, Durham University has developed a programme supporting young people in becoming Science Ambassadors for their own communities (https://www.dur.ac.uk/science.outreach/national/ambassadors/), which is adapted depending upon the needs and interests of the young people. Throughout the year-long programmes, teams of young people work together with community, university and business partners developing their own plans to work with and inspire their peers and community. This has proved successful, even in some of the most deprived areas in England. Ambassadors grow particularly in confidence as well as slightly gaining more enjoyment from presenting. While no statistically significant rises in the numbers stating that they could be a scientist in the future have occurred, schools do report increased parental engagement in Science Ambassador run activities. Further work is currently underway to analyse the longer-term impact on the young people and their communities.

Urban Advantage is a formal-informal partnership between eight informal science learning institutions in New York City and the NYC Department of Education. It works with middle school students (ages 11-14) and focuses on hands-on science inquiry, supported via CPD for teachers and classroom resources, as well as visits to informal science institutions and family workshops. Recent evaluations (e.g. Weinstein et al., 2014) found that participation in Urban Advantage was associated with higher attainment in science (especially among disadvantaged pupils) and increased student confidence in science (aspirations were not measured). Although Urban Advantage is not a UK-based initiative, we feel that it still has relevance to our overall point about the value of longer-term engagements. Additionally, there are limited examples in the UK (or elsewhere) of similar such initiatives, and this scarcity highlights the challenges involved in implementing them.

The Science Capital Teaching Approach (Godec et al., 2017) was developed out of the Enterprising Science project in the UK. It takes a social justice approach to science teaching and is based on a foundation of broadening what counts (whose voices and experiences matter and have a place) in the science classroom. This approach was co-developed with 43 secondary school science teachers over four years. It is a CPD model, involving training teachers in the approach via Saturday sessions and ongoing classroom support during the year. Importantly, it was implemented in science classrooms, meaning that students experienced the approach in an ongoing fashion, over the course of a school year. Research on the approach found a statistically significant increase in the science capital score of young people whose teachers had participated in the project, as well as an increase in aspirations to study science post-16 and improved attitudes to science. Teachers also reported improved behaviour and attainment (L. Archer et al., 2018).

Another CPD programme, Thinking, Talking, Doing Science is run by Science Oxford (https://scienceoxford.com/thinking-doing-talking-science/) and works with primary school teachers over four days across the school year. It aims to support teachers in developing creative science lessons that challenge their pupils to use higher order thinking skills and emphasises the role of discussion in investigations and problem-solving. Evaluations of the programme found benefits to pupils' attainment, as well as attitudes, with particular benefits for girls, previously low-attaining pupils and those eligible for free school meals. As with the Science Capital Teaching Approach, evaluations of this programme highlight the benefits of experiences that extend over the course of a school year.

The Curiosity programme, led by the Wellcome Trust and Children in Need, takes a different approach by exploring the role of science in youth club work outside of schools to target and support young people from disadvantaged backgrounds. Importantly, the programme focuses on key outcomes such as strong self-belief, skills development, positive relationships and emotional wellness rather than science aspirations in their own right. The pilot evaluation, involving 32 different repeated-intervention projects over several months across the UK, noted not only significant reported changes in these outcomes but also had unexpected positive outcomes which included science-related career aspirations (Bright Purpose, 2019). The full-scale programme is currently underway and includes, for example, a programme by youth organisations People Know How and Science Ceilidh in Edinburgh, building on their pilot by running up to 30 youth club sessions annually with the same small group of "at-risk" young people over three years. As well as focusing on supporting confidence, the programme will aim to capture changes in views around science, relationships and determine any potential positive impacts on the young people's transition from primary to high school.

Finally, a number of protracted programmes for secondary and sixth form students to experience undertaking cutting-edge scientific research have emerged in recent years. These have the potential to address the disconnect between science education and what professional scientists actually do (Braund and Reiss, 2006; Yeoman et al., 2017), which may have a role to play in raising, maintaining, or confirming students' aspirations towards STEM subjects (L. Archer et al., 2020; M.O. Archer et al., 2020). Unfortunately, such programmes are rare globally, with those that do exist often lacking active support from researchers and/or targeting of disadvantaged groups (Bennett et al., 2018; M.O. Archer, 2020). One example which avoids both of these issues is the PRiSE (Physics Research in School Environments, http://www.qmul.ac.uk/spa/researchinschools) programme at Queen Mary University of London - a scalable framework for 6-month-long independent research projects based on current physics research topics. PRiSE is supported throughout by active researchers and has equitably involved diverse groups of 14-18 year-old London students (M.O. Archer, 2020; M.O.

Archer et al., 2020). While in some regards this may be seen as similar to the short series, since each project only has a limited number of interventions between researchers and students, the key difference is that there is significant activity between these interventions where students and their teachers work on these research projects. Therefore, the interventions can be thought of as more of a support mechanism for a wider programme, empowering the students to complete their investigations and present them at dedicated student conferences to other PRiSE students, teachers, peers, family, and friends. Longitudinal evaluation (M.O. Archer and DeWitt, 2020) at the 6-month and 3-year stages has shown lasting effects on students' STEM-related confidence and skills. Furthermore, there is evidence suggesting participating students are more likely to continue with physics than beforehand (and to some extent STEM more broadly) as a direct result, with longitudinal evidence 3 years later also revealing greater uptake (to a statistically significant level despite the small sample size) from PRiSE students of both physics and STEM subjects at university than would be expected of physics students nationally. Beyond simply supporting students, PRiSE has been developing teachers' professional practice and even affecting the profile of STEM across their schools (M.O. Archer and DeWitt, 2020). These wider impacts are further reinforced by the significant repeated buy-in from participating schools over several years (M.O. Archer, 2020).

All these programmes share a number of characteristics that likely contribute to their impacts. They are long term - over the course of a school year or more. They involve frequent and repeated activities which students encounter as part of their ordinary classroom learning experience. They both work with and support key influencers in the students' lives, e.g. teachers are leading regular activities. The activities of the interventions are closely integrated with the curriculum in terms of subject content and skills development. They also link together different facets of learners' experience, both inside and outside of the classroom. In understanding, however, how these different aspects contribute to impact upon students, we need to consider the wider picture rather than simply the programmes themselves.

## The big picture

While it appears that long-term interventions have the ability to measurably influence young people's aspirations and identification with science, such efforts are beyond the scope of many organisations. Such repeated-interventions are also challenging to develop, deliver and maintain in settings outside of school, simply because young people generally are not required to be there in the same way that school attendance is mandatory. Thus, it becomes important to consider how to maximise potential impacts of the many activities that can and do take place in formal and informal settings: one-offs, short series or even deeper programmes of engagement. This is where the notion of learning ecosystems or ecologies can be helpful (e.g. Brofenbrenner, 1979). A learning ecology is the context (physical, social, cultural) in which learning takes place. Ecological perspectives on learning acknowledge that young people (and adults, for that matter) learn across a range of contexts, as well as over time, and all of this is influenced in a complex way by previous experiences, background factors, and what follows any given experience. An example diagram of a learning ecology model is shown in Figure 1, where the individual lies at the centre and the layers surrounding have decreasing amounts of direct interaction/influence with them. Such models serve to remind us that our interventions - whether one-offs, short series, or longer, in whatever setting - do not operate in isolation. Moreover, they can be more powerful as learning experiences by linking to other elements in the learning ecosystem. That is, by making a conscious and concerted effort to form links among organisations, young people may be guided towards other experiences that reinforce what they have learned in our activities (Bevan, 2016; Traphagen and Traill, 2014). Doing so can also help us refine our efforts - define our niche - so that it better fits what is needed. In the United States, such initiatives are gaining momentum - and, indeed, a STEM Funders Network has formed to support such efforts in a concerted way (https://stemecosystems.org/about-the-stem-funders-network/).

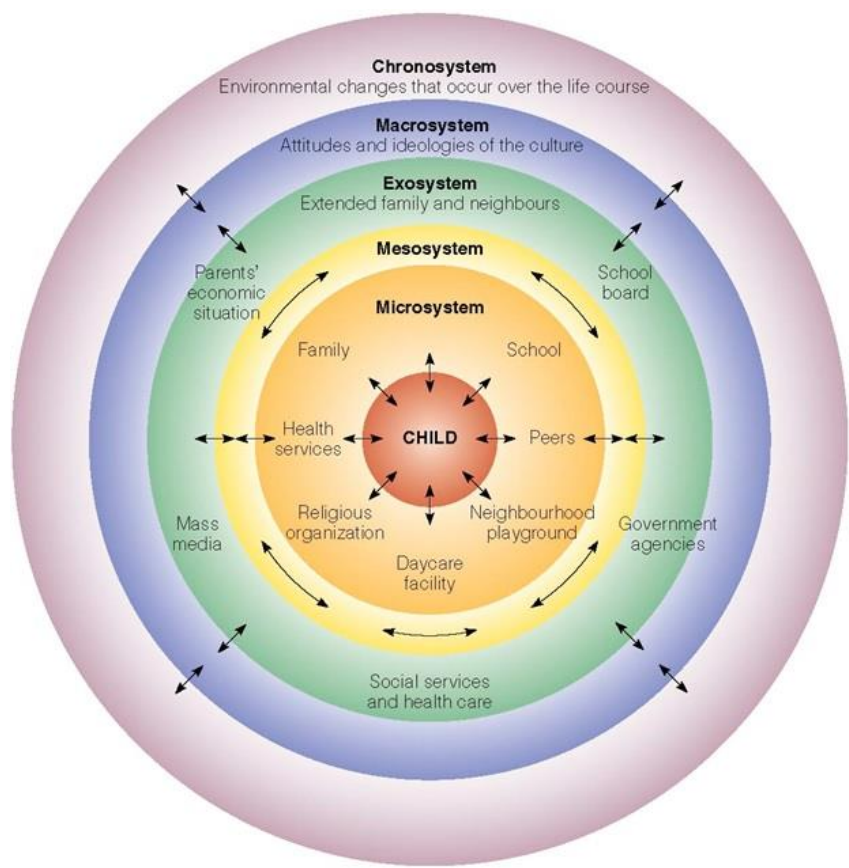

Figure 1: The learning ecology model of Brofenbrenner (1979). From Rhodes (2013).

It is perhaps tempting for organisations that want to effect behaviour or aspiration change to develop an intervention that is based on 'common-sense', but without considering the learning ecology that the child or young person is situated within, without identifying a realistic way in which the desired change could be achieved by the intervention, and with no research base underpinning the intervention. One way to overcome this temptation is to use a Theory of Change approach. This was initially developed to evaluate complex initiatives (Sullivan and Stewart, 2006) and support long-term behaviour change. A key benefit of using a Theory of Change is that it starts with the end goal, and then requires the identification of intermediate outcomes through a process of backwards mapping that will, over time, lead to that goal. In this way causal chains can be identified that link an intervention with evidence-based steps that should eventually lead to the desired change (Davenport et al., 2020). This process is an iterative one, and allows the production of a Theory of Change diagram, such as that shown in Figure 2, which depicts the changes involved in achieving the goal of an intervention.

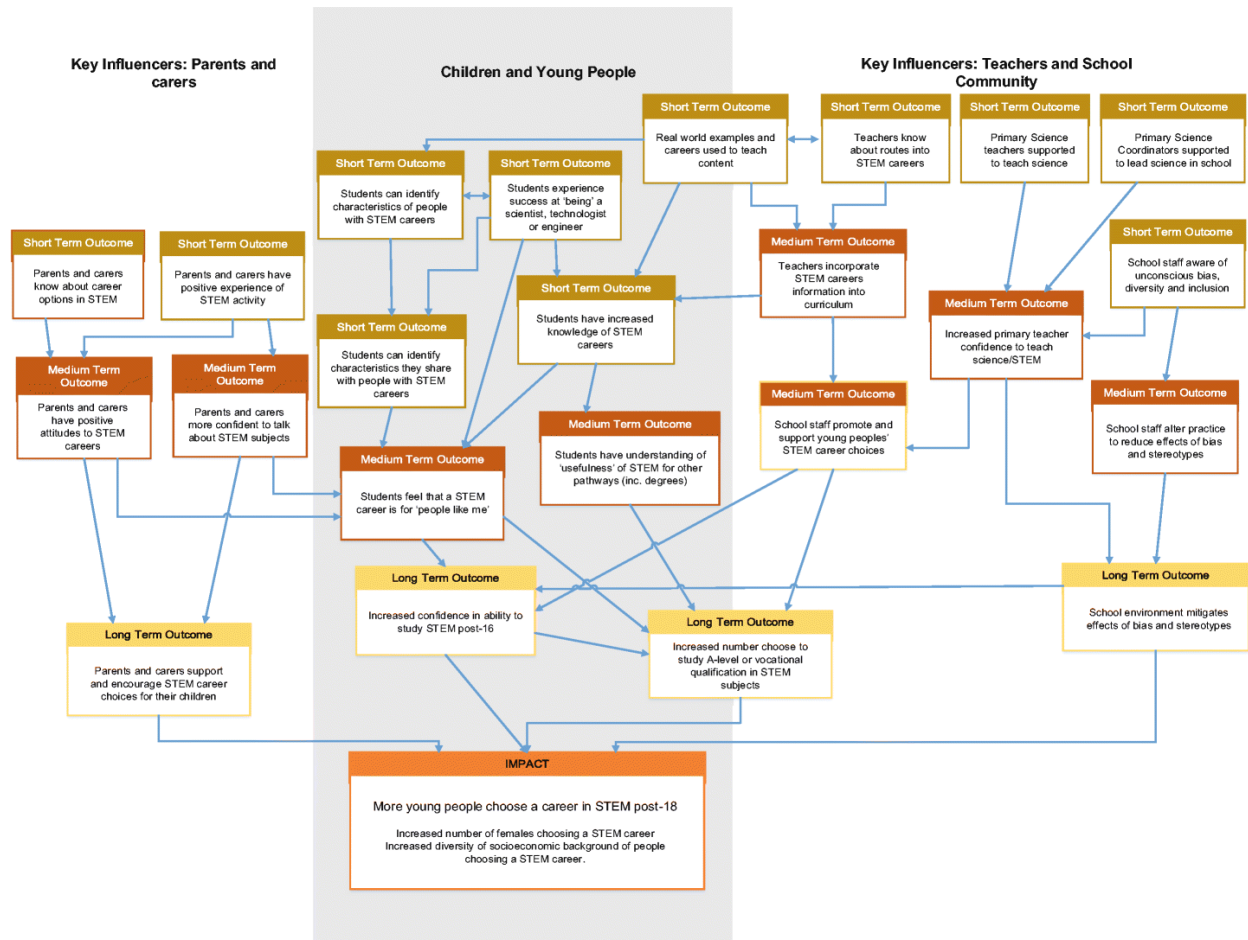

Figure 2: An example Theory of Change diagram for increasing the number of young people choosing a career in STEM post-18. From Davenport et al. (2020).

## Challenges to the sector

The recent review by Robinson and Salvestrini (2020) concerning programmes of engagement with young people advocated for further research to disentangle their individual components in order to better understand the most impactful elements. However, this is perhaps too reductive an approach to developing engagement programmes, given the context of learning ecologies and Theories of Change for impacting on young people's STEM aspirations. Indeed, we have presented several examples demonstrating that the aims of our engagements require considerable effort beyond that which is generally possible at present, and even then they remain complicated and difficult to realise. Therefore, there are many challenges for the sector even with the emergence of these deeper programmes of engagement.

A common theme in many of the repeat interventions programmes discussed is the difficulty in evaluating what impacts, particularly STEM aspirations, the interventions are having. Often changes may be subtle and without the right types of quantitative and/or qualitative analysis, it can appear as if no impact is occurring . Furthermore, longitudinal data is inherently difficult to obtain and the few examples mentioned have had rather small sample sizes. Within universities the growing adoption of the Higher Education Access Tracker (HEAT, [https://heat.ac.uk/](https://heat.ac.uk/)) may help in this regard, however, this alone is not sufficient. For example, HEAT cannot demonstrate a causal link between interventions and outcomes, also it does not provide a broader picture of the young person's entire learning ecology, i.e. what other interventions have they also accessed. These pose real challenges to the engagement sector especially with the current focus on evaluating what impact specific projects have had, such as through REF Impact Case Studies (NCCPE, 2017).

The Theory of Change shown in Figure 2 identifies three groups of stakeholders that are key to increasing the number of young people choosing a STEM career: families, teachers (schools) as well as the children and young people themselves. However, it is noticeable that there is another set of stakeholder groups that do not appear on the diagram: companies, industrial sector bodies, and other STEM organisations. Realistically, there is another Theory of Change that should be developed to explore the organisational changes that need to take place within companies to make them appealing places to work, and address the retention issues that mean a (not insignificant) proportion of young people who enter the STEM workforce do not remain there (Smith & White 2019). The STEM engagement sector (including university outreach organisations) can be seen as 'working' indirectly for these additional stakeholders and doing some of their pre-recruitment work for them. This could lead to a certain moral ambiguity if young people are being encouraged to enter sectors that are (still) not always welcoming to them.

Another issue that is salient in the Theory of Change in Figure 2 is that the learning ecology is extremely challenging to affect. Many of the interventions in STEM engagement continue to focus on the young person, particularly around the points of 'choice' between GCSEs and A-levels. Given the relatively short-term nature of industrial and academic planning where five years is considered long term this is not surprising. There has been some shift in age downwards with recommendations for organisations to work in primary schools - albeit at the upper end of the age range (HEFCE and OFFA, 2013). However, while some interventions involve teachers, very few include parents or carers. In many respects this is understandable, parents of teenagers can be difficult to involve, particularly as some young people may resist some forms of parental involvement in their education (Deslandes and Cloutier, 2002). As young people progress through the education system it becomes more likely that parents and carers will work during the day, and have limited capacity to attend evening events. There has not been, as yet, concerted sector-wide effort to identify how best to engage and include parents in STEM engagement, though programmes like Wellcome Trust and Children in Need's youth work based Curiosity, the STFC funded Association for Science and Discovery Centre's community/family-based Explore Your Universe Phase 4 (https://www.sciencecentres.org.uk/projects/explore-your-universe/phase-4/) and the Scottish

Government's STEM Strategy (2017) focus on Community Learning and Development show steps in this direction.

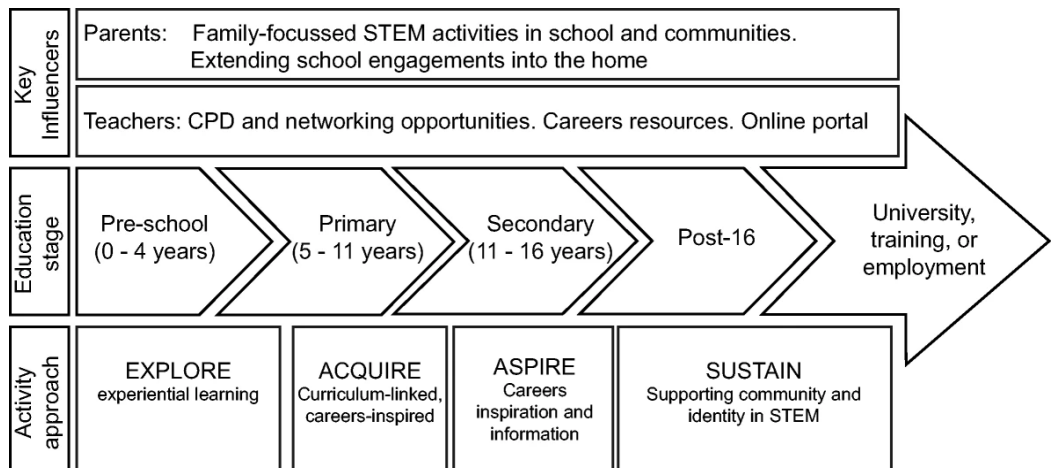

Figure 3: Diagram summarising the educational journey of a child, their key influencers, and the changing nature of activities required during that time. From Davenport et al. (2020).

It is clear that different aims, messages, and activities are needed as students progress through their education. Figure 3 shows a potential way of characterising these throughout their educational journey. However, this raises questions as to who is best placed to be delivering these evolving messages. For example, it is arguable that a university academic is perhaps not the most effective person in discussing (or being a suitable role model for) the wealth of careers beyond academia that continued STEM education can provide. Similarly, many science communicators and engagement professionals do not have the depth of research knowledge or expertise that might be able to sustain a young person's interest through involving students in cutting-edge STEM. This highlights that we need to work more collaboratively, forming partnerships across all those who engage young people, in order to identify who is best placed to deliver the very different types of engagement necessitated at the various points along a young person's educational journey. Such wider collaborative working could, hopefully, shift the focus away from solely the impacts that individual programmes/organisations are having on STEM aspirations and rather promote thinking about how (via Theories of Change and improved evaluations) we can collectively have positive impacts upon young people's learning ecologies.

Wider collaboration, however, comes with practical challenges on who should be managing, coordinating, resourcing, and owning the STEM engagement space. It also comes with data sharing and protection issues regarding the evaluation of the entire learning ecology, rather than simply individual interventions or programmes. It is not currently clear how wide this might need to go. We have presented examples from networks of university departments (i.e. the South East Physics Network) which still suffer from these sector-wide challenges, but whether that means that individual learned societies representing specific STEM subjects, UK Research and Innovation (UKRI) as the umbrella of the UK research councils, or indeed the Government's Department for Education (DfE) needs to step up and take ownership of this mission is up for debate. In Scotland, the Scottish Public Engagement Network (ScotPEN, https://www.scotpen.org/) is beginning to develop a broader network model across universities, learned societies and other informal science organisations and working strategically including locally distributing public engagement funding for Wellcome Trust funded researchers along with working groups to support more impactful work across the ecology.

The UK more widely, could also draw upon models from other countries. For example, in the USA there exists the aforementioned STEM Funders Network along with the Center for Advancement of Informal Science Education (CAISE, https://www.informalscience.org/) - a National Science Foundation funded resource center designed to support and connect the informal STEM education community in museums, media, public programmes and a growing variety of learning environments. Similar public engagement networks exist across Europe, such as Wissenshaft Im Dialog (Science In Dialogue) and Vetenskap & Allmänhet (VA: Public & Science), which aim and strategically support public engagement across stakeholders including universities, learned societies, funders and other informal science organisations across Germany and Sweden respectively.

## Conclusions

Raising and maintaining STEM aspirations are a typical aim of universities' and other practitioners' and organisations' engagements with young people. Aspirations are, however, incredibly complex and difficult to affect, so it is not surprising that any single one-off intervention in isolation, while able to achieve a number of worthy positive outcomes, is unlikely to fundamentally change a young person's future trajectory. We have presented a landscape review and commentary of different emerging programmes of repeated STEM interventions with young people which therefore aim to maximise the likelihood of raising and supporting young people's STEM aspirations. While short series of interventions seem to suffer some of the same limitations as one-offs at present, there is emerging evidence that deeper programmes over the course of months to years that interact not only with the young person but also key components within their wider learning ecology are able to measurably impact on STEM aspirations. However, there is no one-size-fits-all solution to raising and subsequently maintaining the STEM pipeline and a variety of different approaches are required throughout a young person's educational journey.

While we believe that having more deeper programmes of STEM engagement would be the most effective way of positively affecting the STEM landscape overall, we realise that this may not be possible for all types of informal STEM engagement practitioners and organisations. One-offs do, however, have inherent value and might be able to collectively within the wider learning ecology contribute towards building and maintaining aspirations through various different interventions, so long as they present coherent messages that reinforce one another. Programmes of repeated interventions may simply enable providers to control that longer-term messaging, giving individual providers more influence in a consistent way over a great stretch of time than would otherwise be possible. They also inherently allow a greater ability to capture the impacts that the interventions have had. Nonetheless, we advocate that those who develop any STEM engagement activities or programmes consider adopting a Theory of Change approach to critically consider whether what they deliver, be they protracted programmes or one-offs, are likely to achieve (or even just contribute towards) the intended aims. Furthermore, we urge organisations where possible to try to involve and influence the wider ecology in an audience-focused way with every activity to increase the likelihood of any impacts lasting. In lieu of a more joined-up collaborative approach across the entire sector, these considerations at the very least should help us all to make more of a difference.

## Notes on contributors

**Dr Martin Archer** is a UKRI Stephen Hawking Fellow in space physics and public engagement at Imperial College London. He has been a leader in developing award-winning innovative and impactful research-based outreach and public engagement activities for over a decade, sharing the excitement and importance of physics in accessible ways to a variety of often underserved audiences.

**Dr Jennifer DeWitt** is a Senior Research Fellow at the UCL Institute of Education, where she is a member of the team developing and applying the concept of science capital. She is also a research and evaluation consultant, specialising in science learning and engagement, with particular interests in learning in informal settings, including the implications of science capital research for equitable practice in these settings.

**Dr Carol Davenport** is a Senior Lecturer and Director of NUSTEM at Northumbria University, Newcastle. Her research interests include the impact of STEM engagement activities on young people's career choices, the implementation of the Gatsby career benchmarks in subject classrooms and the effects of unconscious bias in primary schools.

**Dr Olivia Keenan** is Director of Outreach and Public Engagement at the South East Physics network. She leads the network's outreach programme and public engagement work. SEPnet work with schools to improve accessibility to, and uptake of, physics. They support their partner universities to engage diverse publics on the research they are conducting. Olivia is passionate about equality and representation in STEM and enjoys working on projects which help embed social justice by removing barriers to access.

**Dr Lorraine Coghill** is Science Outreach Coordinator and NE Regional Representative for Durham University and the Ogden Trust. Lorraine creates, develops and cultivates opportunities for people to work together to explore, enquire and engage others. With nearly 20 years engagement experience, she develops and manages creative and award-winning programmes, and is particularly passionate about training and supporting others in advancing their own engagement practice and confidence.

**Anna Christodoulou** is Collaborative Outreach Officer for Make Happen, the Uni Connect Partnership in Essex. She is working with schools, HE and FE Institutions and charities to raise aspirations and widen participation in higher education throughout Essex. She is a very experienced project manager specialising in outreach and public engagement with research, and an awarded science communicator.

**Samantha Durbin** has been Clothworkers' Associate in Mathematics at the Royal Institution since 2012, managing the Ri's Secondary Mathematics Masterclass network across the UK.. She also co-organises Talking Maths in Public, a skills sharing network and biennial conference for people working in maths outreach/communication.. Samantha was previously a STEM Clubs Advisor at the British Science Association, having completed an MSc in Science Communication at the University of the West of England in 2012 and Master of Mathematics at the University of Bath in 2010.

**Dr Heather Campbell** completed an Mphys in Astrophysics with Research Placement at the University for Sussex in 2009. Then, a PhD on the explosions of stars, Type Ia Supernovae to explore the accelerated expansion of the Universe, at the University of Portsmouth. Heather worked as a postdoctoral researcher on the Gaia Satellite at the University of Cambridge. For the last 4 years Heather has been the SEPnet/Ogden Public Engagement and Outreach Manager at the University of Surrey.

**Lewis Hou** is an interdisciplinary education and cultural participation specialist. He is particularly interested in public and community engagement approaches that build meaningful and equitable relationships with diverse groups beyond the "already converted" and was the recipient of the Royal Society of Edinburgh's Public Engagement Innovator Medal in 2018. He founded and directs the Science Ceilidh, an education and community organisation exploring curiosity, creativity, equity, and health and wellbeing with adults, youth groups and schools across Scotland.